\documentclass{article}

\usepackage{arxiv}

\usepackage[utf8]{inputenc} 
\usepackage[T1]{fontenc}    
\usepackage{hyperref}       
\usepackage{url}            
\usepackage{booktabs}       
\usepackage{amsfonts}       
\usepackage{nicefrac}       
\usepackage{microtype}      
\usepackage{bm}
\usepackage{amssymb}
\usepackage{graphicx}
\usepackage{epstopdf}
\usepackage{algorithm}
\usepackage{algpseudocode}
\usepackage[round]{natbib}
\usepackage{placeins}
\usepackage{float}
\usepackage{pagecolor}

\usepackage{mathtools}

\title{Generating Artificial Core Users for Interpretable Condensed Data}

\author{
  Amy ~Nesky \\
  Computer Science and Engineering\\
  University of Michigan, Ann Arbor\\
  Ann Arbor, MI 48109\\
  \texttt{anesky@umich.edu} \\
   \And
 Quentin F.~Stout \\
  Computer Science and Engineering\\
  University of Michigan, Ann Arbor\\
  Ann Arbor, MI 48109\\
  \texttt{qstout@umich.edu} \\
}

\begin{document}
\maketitle
\pagecolor{white}

\newcommand{\XX}[1]{}      

\begin{abstract} 

Recent work has shown that in a dataset of user ratings on items there exists a group of Core Users who hold most of the information necessary for recommendation. This set of Core Users can be as small as 20 percent of the users. 
Core Users can be used to make predictions for out-of-sample users without much additional work. Since Core Users substantially shrink a ratings dataset without much loss of information, they can be used to improve recommendation efficiency. 
We propose a method, combining latent factor models, ensemble boosting and K-means clustering, to generate a small set of Artificial Core Users (ACUs) from real Core User data. Our ACUs have dense rating information, and improve the recommendation performance of real Core Users while remaining interpretable.

\XX{
can be used to enhance Core Users?
}


\end{abstract}

\keywords{Recommender Systems \and Core Users \and Synthetic Data Generation}

\section{Introduction}

Recommendations Systems improve the user experience by providing personalized, curated recommendations in a large and complex information space.  Collaborative Filtering methods exploit community data to uncover correlations between users and items that can be used to make recommendations.  Within collaborative filtering, latent factor models are considered state-of-the-art; these models approximate highly redundant data with low rank matrix decompositions \citep{MF,MF1,MF2,MF3,textBook}. 

Unfortunately, many matrix factorizations methods are transductive and cannot be leveraged to make predictions for users outside of the original training set  \citep{textBook}. Orthonormal matrix factors can be used more easily to make out-of-sample predictions because the matrix factors capture geometric traits of the item and user spaces.  However, obtaining orthonormal factors can be expensive for large datasets. 

The majority of research in recommender systems focuses on developing new collaborative filtering and hybrid methods, which are often highly optimized for a specific application. There is a smaller body of research looking at identifying the most useful users who carry most of the relevant information, and separating out those users as Core Users for making recommendations.  With a smaller core set of users, making out-of-sample predictions becomes a reasonable task. Reducing a dataset size without much loss of information improves recommendation efficiency as well as storage costs; numerous fields outside of recommender systems would benefit from this ability. 

However, available Core User methods have limited representation ability, in that they are selected from existing user data. In other words, the recommendation success of Core Users is bounded by quality of user data available. 
Improving the recommendation accuracy of Core Users makes data abstraction more effective for applications in data augmentation, bots mimicking population behavior, data mining, privacy, statistics and many more. In this paper, we develop a method of generating Artificial Core Users (ACUs) that improves the recommendation accuracy of real Core Users. 
We combine latent factor models, ensemble boosting and K-means clustering, to generate a small set of Artificial Core Users (ACUs) from real Core User data. Our ACUs incur a small amount of additional memory storage when compared to real Core Users, but remain a reduction in memory storage compared to the original dataset. Artificial Core Users improve the recommendation accuracy of real Core Users while remaining good centroids for the complete recommendation dataset. Since ACUs act as good centroids for the complete dataset, ACUs blend in well with the real dataset even though they are generated artificially. But unlike real Core Users, ACUs have complete ratings on all items, providing more immediately interpretable information to scientists.

\XX{

There are numerous applications of finding a small representative set outside of user recommendations, including   

other people who might care about this: statisticians, data mining, privacy??

}


\section{Related Work}

The inductive matrix completion problem assumes that a ratings matrix is generated by applying feature vectors to a known low-rank matrix  \citep{InductiveMatrixCompletion,InductiveMatrixCompletion2}.  This is relevant for making out-of-sample recommendations if we assume that the latent factors contain some ground truth about the dataset that applies to out-of-sample users. As mentioned above, latent factors produced by most methods are sadly transductive  \citep{textBook}. 
To combat this, some have worked on improving the efficiency of using a singular value decomposition in recommender systems \citep{SVD}.

Using clustering is one of the earliest attempts to decrease the number of users needed to make recommendations \citep{textBook}; rating predictions can be made using only information from the relevant cluster. Alternatively, \cite{CoreUsers} study the relevance of different users and find that there exists an \textquoteleft \textquoteleft information core" made up of some key users. They found that the number of the Core Users is around $20\%$ of the entire dataset, and that the recommendation accuracy produced by only relying on the Core Users can reach 90 percent of that produced using every user in the dataset. \cite{CoreUsers} use a generalized K-nearest Neighbor algorithm using various relevancy metrics to measure \textquoteleft nearness'. They ran experiments using degree-based, resource-based and similarity-based measures, to select their Core Users; all of these measures are graphical in nature. Since this work a few other methods for selecting Core Users have emerged. \cite{Constructing} use a long-tail-distribution-based measure to select their core uses. \cite{CoreUsers2} introduced a new measure to identify Core Users based on trust relationships and interest similarity; this work extends beyond graphical knowledge to include the semantic meaning of items.  \cite{CoreUsers3} use a combination of the measures proposed in \citep{CoreUsers2} and \citep{Constructing}. 

\XX{
talk about how our algebraic method is more scalable than there graphical methods
talk about the difference in the number of users needed
talk about how we're able to produce dense ratings for a smaller number of users
}

Recently, deep learning has made an appearance in recommender systems either in the form of integration models or neural network models \citep{DLRSsurvery}. 
Integration models use neural networks to uncover features in auxiliary information, like item descriptions \citep{DLRSAI1,DLRSAI2}, user profiles \citep{DLRSAI} and knowledge bases \citep{DLRSAI3}. The uncovered features are then incorporated into a collaborative filtering framework to produce hybrid recommendations. 
Neural network models on the other hand perform collaborative filtering directly via modeling the interaction function between users and items \citep{DLRS,DLRS1,DLRS2,DLRS3,Autoencoder}. Using deep learning to make recommendations means the models are better equipped to recognize nonlinear relationships in the data, but it also means that the models inherit all the training difficulties neural networks face compounded by the difficulties of working with sparse data.

There is a small body of work that injects fake users into a recommendation system either for adversarial goals \citep{AdversarialFakes,AdversarialFakes1,AdversarialFakes2} or utilitarian data augmentation goals \citep{FilterBots}. Typically, the fake users are hand-coded, but \citep{AdversarialFakes2} create adversarial fake user profiles for a recommendation system using generative adversarial nets.  \cite{FilterBots} on the other hand used simple content filtering bots to generate a few users with dense ratings to improve the recommendations. 
Tackling the new user, or cold start, problem is an issue in recommender systems that has brought about some work on determining which item preferences and item features are most informative \citep{InfoTheory,InfoTheory2,newUser}.

\XX{

numerous fields outside of recommender systems would benefit from the ability to reduce their dataset into a small representative set without much loss of information. 
other people who might care about this: statisticians, data mining, privacy??

}


\section{Methodology}

We propose an offline technique to generate a small set of Artificial Core Users (ACUs) who's dense ratings matrix can be stored in condensed, low order form and used to make recommendations for out-of-sample users without much additional work. This method uses latent factor models, ensemble boosting and K-means clustering to learn a small group of artificial users who's feature vectors capture correlations in the full dataset. This is a collaborative filtering method that uses an existing sparse ratings matrix for a set of users and items.
 
The algorithm requires a set of training users represented by their ratings on a given set of items, which will be used to update ACU ratings during learning. To measure the abstraction of the ACUs, we split the dataset into two mutually exclusive sets: testing users and  training users. Let $R$ be an $m\times n$ sparse ratings matrix for $m$ training users and $n$ items.\footnote{We did choose to normalize the variance and mean center the rows of $R$ before proceeding; any adjustments made here can be accounted for at recommendation time.} 
The ACU set size will be small relative to the size of the dataset; if $s$ is the number of ACUs, we pick $s$ such that $s\ll m$.  Let $R_{ACU}$ denote the $s\times n$ ratings matrix for our ACUs. There are a number of ways one could initialize the ACU ratings;
in our work, ACU ratings will be initialized from Core User ratings.
\footnote{As with any learning process, one can see a lot of improvement by selecting the right initialization. There is room to try supplementing Core User ratings to improve the initialization of $R_{ACU}$.}

The training algorithm is given in Algorithm \ref{algGenerateACUs}. We trained row blocks of $R_{ACU}$ together using batches of training users. 
By learning ACUs in blocks, we incorporate boosting results and reduce our workload \citep{Schapire90thestrength,boosting}. Algorithm \ref{algGenerateACUs} learns orthonormal decompositions for each block, leveraging the relevance of orthonormal decompositions to out-of-sample relationships.\footnote{Computing the singular value decomposition of a large matrix is known to be time consuming, so learning in blocks saves quite a bit of time.}
Let $R_{ACU}^{(b)}$ denote the $b^{th}$ row block in $R_{ACU}$, and $R_{i}$ denote the sparse ratings matrix of the $i^{th}$  batch of training users.  For simplicity, we assume that each block of $R_{ACU}$ has the same number of rows and similarly each training batch has the same number of users. We divide $R_{ACU}$ into $\zeta$ equal sized row blocks such that for $b\in\{0,\dots,\zeta\}$ $R_{ACU}^{(b)}$ is a $\left(\frac{s}{\zeta}\times n\right)$ matrix. Let $I$ denote the training batch index set such that for $i\in I$, $R_{i}$  is a $\left(\frac{m}{|I|}\times n\right)$ matrix. The variables $\alpha$ and $\beta$ are learning rates, while the variables  $\lambda$ and $\gamma$ are regularization coefficients. Lines \ref{gd1} and \ref{gd2} in Algorithm \ref{algGenerateACUs} are derived from stochastic gradient descent algorithms. Line \ref{km} updates the rows of $U_{ACU}$ as the $\frac{s}{\zeta}$-means of $\frac{m}{|I|}$ row vectors in $U_i$, which is described in detail in Algorithm \ref{algMeansUpdate}. 
Embedding Algorithm \ref{algMeansUpdate} inside Algorithm \ref{algGenerateACUs} effectively trains the Artificial Core User user-space matrix, $U_{ACU}$, with Mini-Batch K-means \citep{mbkm}. By incorporating Algorithm \ref{algMeansUpdate} into the learning process we expect to keep the resulting Artificial Core User ratings matrix interpretable. That is, we expect the resulting Artificial Core User ratings to resemble that of real users, but provide complete rating information in place of sparse data. 
As with any learning method, the amount of regularization will be problem dependent; one may find that they can get away with $\lambda,\gamma=0$ because lines \ref{optional} and \ref{svd1}-\ref{svd2} in Algorithm \ref{algGenerateACUs} have a regularizing effect. 
Lines \ref{svd1}-\ref{svd2} ensure that we don't stray to far from an orthonormal decomposition during learning. 
The while loop on line \ref{innerwhile} takes a simplistic approach to the inductive matrix completion problem given a set of features; this subprocess could likely be improved by existing work \cite{InductiveMatrixCompletion,InductiveMatrixCompletion2}. 

Algorithm \ref{algGenerateACUs} should be run offline to produce a set of Artificial Core Users. After which, one can use $V_{ACU}S_{ACU}$ to make faster recommendations for out-of-sample users. One may also be interested in using $R_{ACU}$ as a condensed version of their dataset. 

This process resembles the way one would train a neural network and shares similarities with neural network models \citep{DLRS,DLRS1,DLRS2,DLRS3,Autoencoder}.  Neural network models generally use autoencoders. One layer Neural Collaborative Autoencoders with no output activation can be reformulated to parallel matrix factorization \citep{Autoencoder}. Unlike Algorithm \ref{algGenerateACUs}, existing these models are not concerned with interpretability.


\XX{Talk about the relationship between our work and autoencoders}

We evaluate our work using two metrics: item vectors testing error and sparse K-means error. To measure the item vectors testing error, we use Algorithm \ref{algItemVectorTestingError} on an independent set of testing users. 
Algorithm \ref{algItemVectorTestingError} is a simplistic recommender model that tries to predict missing user ratings.
It decomposes the row blocks of $R_{ACU}$ into orthogonal components using a singular value decomposition.  Then, for a set of real testing users, Algorithm \ref{algItemVectorTestingError} optimizes a set of unit vectors to accompany the item vectors extracted from the decomposition of $R_{ACU}$ - there is no new learning done in the item vectors.  In this way, Algorithm \ref{algItemVectorTestingError} evaluates the generality of the item vectors extracted from $R_{ACU}$ and the potential for high quality recommendations using a more elaborate learning model. It should be noted that, as a simplistic recommender model, Algorithm \ref{algItemVectorTestingError} produces far from state-of-the-art recommendation results, but is useful for our purposes of evaluating whether Artificial Core Users better capture the information in a recommendation dataset than real Core Users do. 
To compute the sparse K-means error of $R_{ACU}$ we make a few modifications to Algorithm \ref{algMeansUpdate} using a sparse first input matrix; this results in Algorithm \ref{algSparseMeansError}. Algorithm \ref{algSparseMeansError} estimates how well the Artificial Core Users represent an average collection of real users, $R_{ACU}$, by measuring how well the ACUs serve as centroids for our real testing users.
\XX{
\[
U_{ACU}\leftarrow (1 -\beta)U_{ACU} + \beta \mbox{K-means}(U_{ACU},U)
\]
}

\begin{algorithm}[h]
  \caption{Generate\textunderscore ACUs($R$)}\label{algGenerateACUs}
  \begin{algorithmic}[1]
  \State Let $m$ be the number of users (row dimension of $R$) and $n$ be the items (column dimension of $R$). 
      \State Initialize constants $\alpha$, $\lambda$, $\beta$, $\gamma$. Initialize matrix $R_{ACU}$ with dimensions $s\times n$.
      \While{not done}\label{outerwhile}
      \For{$b=0$ to $\zeta$}
      \State Find orthonomal $\left(\frac{s}{\zeta}\times k\right)$ matrix $U_{ACU}^{(b)}$, orthonomal $\left(n\times k\right)$ matrix $V_{ACU}^{(b)}$ and diagonal $(k\times k)$ matrix $S_{ACU}^{(b)}$ such that $U_{ACU}^{(b)} S_{ACU}^{(b)}\left(V_{ACU}^{(b)}\right)^\top \approx R_{ACU}^{(b)}$ and $k\le \min\left(\frac{s}{\zeta}, n\right)$.\label{loworder}
      \State Set $V_{ACU}^{(b)} = V_{ACU}^{(b)}S_{ACU}^{(b)}$.
             \State Select $i$ randomly from training user batch index set $I$.
            \State Initialize $U_i$. 
             \State $j=0$.
       \While{not done} \label{innerwhile}
       \State Compute sparse $E = R_we -  U_we \left(V_{ACU}^{(b)}\right)^\top$ for the specified entries of $R_i$. Other entries of $E$ remain zero.
       \If{not done}
       \If{$j \mod 2 = 0$}
       \State $U_we \leftarrow (1 -\beta\gamma)U_we + \beta E V_{ACU}^{(b)}$\label{gd1}.
       \State Normalize the columns of $U_i$\label{optional}.
	\Else
	\If{$j \mod 4 = 1$}
       \State Means\textunderscore Update$\left(U_i,U_{ACU}^{(b)}\right)$\label{km}. 
	\Else
        \State $V_{ACU}^{(b)}\leftarrow (1 -\alpha\lambda)V_{ACU}^{(b)} + \alpha E^\top U_i$.  \label{gd2}
       \EndIf
       \EndIf
      \State  $R_{ACU}^{(b)} = U_{ACU}^{(b)} \left(V_{ACU}^{(b)}\right)^\top$\label{svd1}.
      \State Find orthonomal matricies $U_{ACU}^{(b)}$, $V_{ACU}^{(b)}$ and diagonal matrix $S_{ACU}^{(b)}$ such that $U_{ACU}^{(b)} S_{ACU}^{(b)}\left(V_{ACU}^{(b)}\right)^\top \approx R_{ACU}^{(b)}$\label{svd2}.
      \State Set $V_{ACU}^{(b)} = V_{ACU}^{(b)}S_{ACU}^{(b)}$.
        \State $j =j+1$.
        \EndIf
       \EndWhile
      \EndFor
      \EndWhile \label{euclidendwhile}
  \end{algorithmic}
\end{algorithm}
\begin{algorithm}[h]
  \caption{Item\textunderscore Vectors\textunderscore Testing\textunderscore Error($R, R_{ACU}, \zeta$)}\label{algItemVectorTestingError}
  \begin{algorithmic}[1]
  \State Let $m$ be the number of users (row dimension of $R$), $n$ be the items (column dimension of $R$ and $R_{ACU}$) and $s$ be the number of ACUs (row dimension of $R_{ACU}$). 
  \State Initialize constants $\beta$, $\gamma$.
      \For{$b=0$ to $\zeta$}
      \State Find orthonomal $\left(\frac{s}{\zeta}\times k\right)$ matrix $U_{ACU}^{(b)}$, orthonomal $\left(n\times k\right)$ matrix $V_{ACU}^{(b)}$ and diagonal $(k\times k)$ matrix $S_{ACU}^{(b)}$ such that $U_{ACU}^{(b)} S_{ACU}^{(b)}\left(V_{ACU}^{(b)}\right)^\top \approx R_{ACU}^{(b)}$ and $k\le \min\left(\frac{s}{\zeta}, n\right)$.\label{loworder}
      \EndFor
      \State Let $V_{ACU}=\left[V_{ACU}^{(0)}\cdots V_{ACU}^{(\zeta)} \right]$.
      \State Aggregate $S_{ACU}^{(b)}$ for $b=0$ to $\zeta$ along diagonal blocks to form the diagonal matrix $S_{ACU}$.
       \State Set $V_{ACU}^{(b)} = V_{ACU}^{(b)}S_{ACU}^{(b)}$.
             \State Randomly select $80\%$ of the nonzero entries in $R$ as training entries and let the other $20\%$ be the probe entries.
             \State Define $R^{(T)}$ as the sparse matrix made up of the training entries of $R$, and define $R^{(P)}$ as the sparse matrix made up of the probe entries of $R$ such that $R= R^{(T)} + R^{(P)}.$
            \State Initialize $U$. 
       \While{not done} \label{while}
       \State Compute sparse $E = R^{(T)}-  U V_{ACU}^\top$ for the specified entries of $R^{(T)}$. Other entries of $E$ remain zero.
       \If{not done}
       \State $U\leftarrow (1 -\beta\gamma)U + \beta E V_{ACU}$\label{gd1}.
       \State Normalize the columns of $U$\label{optional}.
        \EndIf
       \EndWhile
      \State Compute sparse $E = R^{(P)} -  U V_{ACU}^\top$ for the specified entries of $R^{(P)}$.
      \State Return the average absolute value of an entry in the $E$ for the specified entries of $R^{(P)}$.
  \end{algorithmic}
\end{algorithm}

\begin{algorithm}[h]
  \caption{Means\textunderscore Update($A,B$)}\label{algMeansUpdate}
  \begin{algorithmic}[1]
  \State  Initialize lists $L_i$ for $we = 0$ to the row dimension of $B$.
         \For{$r=0$ to the row dimension of $A$}
         \State min\textunderscore val $= \infty$, min\textunderscore index $=0$.
       \For{$l=0$ to the row dimension of $B$}
       \If{The distance between the $r^{th}$ row of $A$ and the $l^{th}$ row of $B$ is less than min\textunderscore val}
       \State min\textunderscore val $=$ this distance.
       \State min\textunderscore index $=l$.
        \EndIf
       \EndFor
       \State $L_{\mbox{min\textunderscore index}}.push(r)$.
        \EndFor
        \For{$l=0$ to the row dimension of $B$}
         \If{ list $L_l$ is non-empty}
         \State Set the $l^{th}$ row of $B$ to the weighted average of the rows of $A$ with indices stored in list $L_l$.
          \EndIf
          \EndFor
    \end{algorithmic}
\end{algorithm}

\FloatBarrier

\begin{algorithm}[h]
  \caption{Sparse\textunderscore Means\textunderscore Error($R,R_{ACU}$)}\label{algSparseMeansError}
  \begin{algorithmic}[1]
  \State avg\textunderscore error = 0.
  \State entry\textunderscore count = 0.
         \For{$r=0$ to the row dimension of $R$}
         \State min\textunderscore val $= \infty$, min\textunderscore index $=0$.
       \For{$l=0$ to the row dimension of $R_{ACU}$}
       \If{The sparse distance between the $r^{th}$ row of $R$ and the $l^{th}$ row of $R_{ACU}$ is less than min\textunderscore val for specified entries of $R$}
       \State min\textunderscore val $=$ this distance.
       \State min\textunderscore index $=l$.
        \EndIf
       \EndFor
       \For{sparse entries in the $r^{th}$ row of $R$}
        \State entry\textunderscore count += 1.
        \State temp\textunderscore err = absolute difference between entry of the $r^{th}$ row of $R$ and the corresponding entry in the min\textunderscore index row of $R_{ACU}$.
        \State avg\textunderscore error += (temp\textunderscore err - avg\textunderscore error) / entry\textunderscore count.
        \EndFor
 \EndFor
 \State Return avg\textunderscore error.
    \end{algorithmic}
\end{algorithm}



\section{Experimental Results}

Our main objective in our experiments is comparability across methods rather than state-of-the-art performance. To make the error on the probe entry sets comparable when testing with Algorithm \ref{algItemVectorTestingError}, we aimed for similar errors on the training set; to accomplish this goal we stopped  the algorithm early when necessary.

All experiments are run on the Bridges' NVIDIA P100 GPUs through the Pittsburgh Supercomputing Center. We ran experiments using the MovieLens ml-20m dataset \citep{MovieLens}. We normalized the variance and mean centered the rows of the dataset ratings matrix before proceeding.\footnote{Any adjustments made here can be accounted for at recommendation time.}

\XX{
talk about learning rates. only allowed to learn for max 500 iterations
}

We will compare our work for generating ACUs to existing methods for selecting Core Users from \citep{CoreUsers,Constructing,CoreUsers2,CoreUsers3}.  All of the methods for finding Core Users tested here are derived from the K-nearest neighbor algorithm.  These methods use some metric to determine the pair-wise similarity between all existing users. Then, for each user, a list of the top-K most similar users is generated; from these combined lists the Core Users are selected. In our Core User experiments, we collect the top-50 most similar users for each user. 
Existing methods differ in their pair-wise similarity metric for users, and in their selection method within the compiled top-K most similar user lists.

Recall that the cosine similarity of two vectors $A$ and $B$ is $(A\cdot B)/(||A||\cdot||B||)$. In all of our Core User experiments, the pair-wise similarity between all existing users will be calculated in one of two ways: it will either be the cosine similarity of the users' ratings vectors, or it will be the cosine similarity of their boolean vectors where their boolean vectors indicate only whether or not an item has been rated - not how well the user liked it.

Once the lists of the top-K most similar users to each user have been generated, one can either count the frequency of a given user's appearance in the top-K lists and take the most frequent users as the Core Users, or one can weight a user's appearance in a list by the inverse of the rank within the list that user appeared; the first way we will refer to as frequency-based and the second way we will refer to as rank-based.

The final variant is whether or not to consider \textquoteleft hidden' ratings in the cosine similarity of users.  Without considering hidden ratings, two users with no overlapping rated items would have zero similarity, but if the users have rated items that are similar to one another this metric seems insufficient. \cite{CoreUsers2,CoreUsers3} consider semantic relationships between items to determine item similarity, here we will use the cosine similarity of the item vectors where the item vectors are the rating data from all of the users for the given item.\footnote{For each item, a list of the top-K most similar items is generated.}  After computing the item similarity, a missing rating may be substituted for with a weighted average of similar rated items, where the item similarity can be used as the weight.

\XX{

\citep{CoreUsers2} select Core Users by measuring the similarity between each pair of users and collecting the users who rank highly similar to many users. In their work, similarity between users is a weighted cosine similarity where the weight is calculated using semantic item similarity. Incorporating item similarity into their calculation allows them to alleviate some of the contsraints caused by the sparsity of the rank matrix. We will compare our work to methods from \citep{CoreUsers2}.

Using clustering is one of the earliest attempts to decrease the number of users you need to work with in order to make recommendations \citep{textBook}; rating predictions can be made using only information from the relevant cluster. Alternatively, in \citep{CoreUsers}, authors study the relevance of different users and find that there exists an \textquoteleft \textquoteleft information core" made up of some key users. They found that the number of the Core Users is around 20 percent of the entire dataset, and that the recommendation accuracy produced by only relying on the Core Users can reach 90 percent of that produced using every user in the dataset. 

\citep{CoreUsers} use a generalized K-nearest Neighbor algorithm using various relevancy metrics to measure \textquoteleft nearness'. They ran experiments using degree-based, resource-based and similarity-based measures, to select their Core Users; all of these measures are graphical in nature. Since this work a few other methods for selecting Core Users have emerged. \citep{Constructing} use a long-tail-distribution-based measure to select their core uses. \citep{CoreUsers2} introduced a new measure to identify Core Users based on trust relationships and interest similarity; this work extends beyond graphical knowledge to include the semantic meaning of items.  \citep{CoreUsers3} use a combination of the measures proposed in \citep{CoreUsers2} and \citep{Constructing}. Our method for generating ACUs is algebraic instead of graphical, so it is more scalable than the methods developed for finding Core Users.  Additionally, the Artificial Core User ratings we produce are dense, so we can store significantly less of them and maintain recommendation accuracy.
}

\subsection{MovieLens}

The MovieLens ml-20m contains ratings for 138493 users on a set of 27278 movies \citep{MovieLens}. This section will discuss the performance of ACUs compared to existing Core User methods using the MovieLens ml-20m dataset.

\begin{table}[h]
\caption{Item Vectors Testing Error of 13000 Core Users collected with previously existing methods using the ml-20m dataset \citep{MovieLens}. We averaged the results of Algorithm \ref{algItemVectorTestingError} over 75 runs where each run was given 200 independent testing users and 2600 of the Core User Item Vectors, or $\approx50\%\ $ of the singular value mass. In each run, we stopped Algorithm \ref{algItemVectorTestingError} when we reached a training error of 0.2. }\label{ItemVectorsTestingErroronCoreUsers}
\begin{center}
\begin{tabular}{cccc|}
\textbf{Item} &\textbf{Ratings} &\textbf{Frequency-} &\textbf{Probe} \\
\textbf{Similarity} &\textbf{Used} &\textbf{Based} &\textbf{Entry} \\
\textbf{Used} & &&\textbf{Set Error} \\
\hline \\
    yes & yes& yes&1.41\\
   yes & yes& no&\bf{1.38}\\
   yes & no& yes&1.71\\
   yes & no& no&1.67\\
     no & yes& yes&1.45\\
   no & yes& no&1.39\\
   no & no& yes&1.73\\
   no & no& no&1.70\\
\end{tabular}
\end{center}
\end{table}

Table \ref{ItemVectorsTestingErroronCoreUsers} shows the performance of the various previously existing Core User selection methods when tested using the item vectors testing error; each method was used to collect 13000 Core Users or a little under $10\%$ of the original MovieLens ml-20m dataset size.  This is half the number of Core Users necessary to maintain recommendation accuracy as claimed in \citep{CoreUsers}, but our aim is comparing the viability of these methods. To test each method we used 13 row blocks, or 1000 users per block.
We found that the performance is sensitive to the number of item vectors retained as latent factors in the line \ref{loworder} of Algorithm \ref{algItemVectorTestingError}; we'll discuss reasons for this a bit further down. 
We chose to use $20\%$ of the singular values, for a total of $2600$ latent factors across all the blocks. 
To make the error on the probe entry sets comparable across methods, we stopped running Algorithm \ref{algItemVectorTestingError} when the error on the training set had reached $0.2$.
The average absolute value of an entry in the ratings matrix, after centering, is $\approx0.8$; in other words, the error when always predicting that a missing rating will be the mean, zero, is $\approx0.8$. Therefore, we ran Algorithm \ref{algItemVectorTestingError} for various Core User methods until the error on the training set was $\approx75\%$ better than simply always guessing the mean.
The first column in Table \ref{ItemVectorsTestingErroronCoreUsers} indicates whether or not hidden ratings taken from the item similarities were incorporated into the calculation of the user similarities. The second column indicates whether the cosine similarity of the user vectors is taken using the actual item ratings or just the item booleans. The third column indicates whether we used a frequency-based or a rank-based approach to select the Core Users from the aggregated lists of the top-50 most similar users to each other user. 
These results support previous literature suggesting that the best method for selecting Core Users considers item similarity, compares ratings rather than booleans, and uses a rank-based selection approach.  
%

%


For all methods used in Table \ref{ItemVectorsTestingErroronCoreUsers}, Algorithm \ref{algSparseMeansError} returns an error of $\approx0.715$; this error is $12\%$ better than the error when using only one centroid with the mean at the origin.

\FloatBarrier

Figures \ref{CUSVfig} and \ref{CUSVdifffig} help to explain why the performance of Algorithm \ref{algItemVectorTestingError} is sensitive to the number of item
vectors retained as latent factors in the line \ref{loworder}. They compare the singular values of the 13000 Core User ratings matrix where the Core Users are selected using the most competitive selection method: the method used in the second row of Table \ref{ItemVectorsTestingErroronCoreUsers} to the singular values of an equivalently sparse matrix with random ratings as the non-zero values. Singular values lay out the behavior of a matrix as a mapping between spaces, and the Marchenko-Pastur theorem \citep{MP}, describes the asymptotic behavior of the singular values of large random matrices.  Figures \ref{CUSVfig} and \ref{CUSVdifffig} show that the singular values of the Core User ratings matrix only significantly differ from those of a random matrix in the lowest order singular values, which are, by convention, the largest singular values. In fact, beyond the first $1\%$ of singular values, the singular values of the Core User ratings matrix differ by at most 3.6 from those of a random matrix, with larger order singular values contributing less influence over the behavior of the matrix. So, while the larger order singular values of the Core User ratings matrix are non-zero, which is generally how we measure relevance, this relationship suggests that the larger order singular vectors may not be informative and may actually make learning more difficult by adding noise.

\begin{figure}[h]
  \centering
  \begin{minipage}[t]{.48\textwidth}
   \centering
  \includegraphics[width=\linewidth]{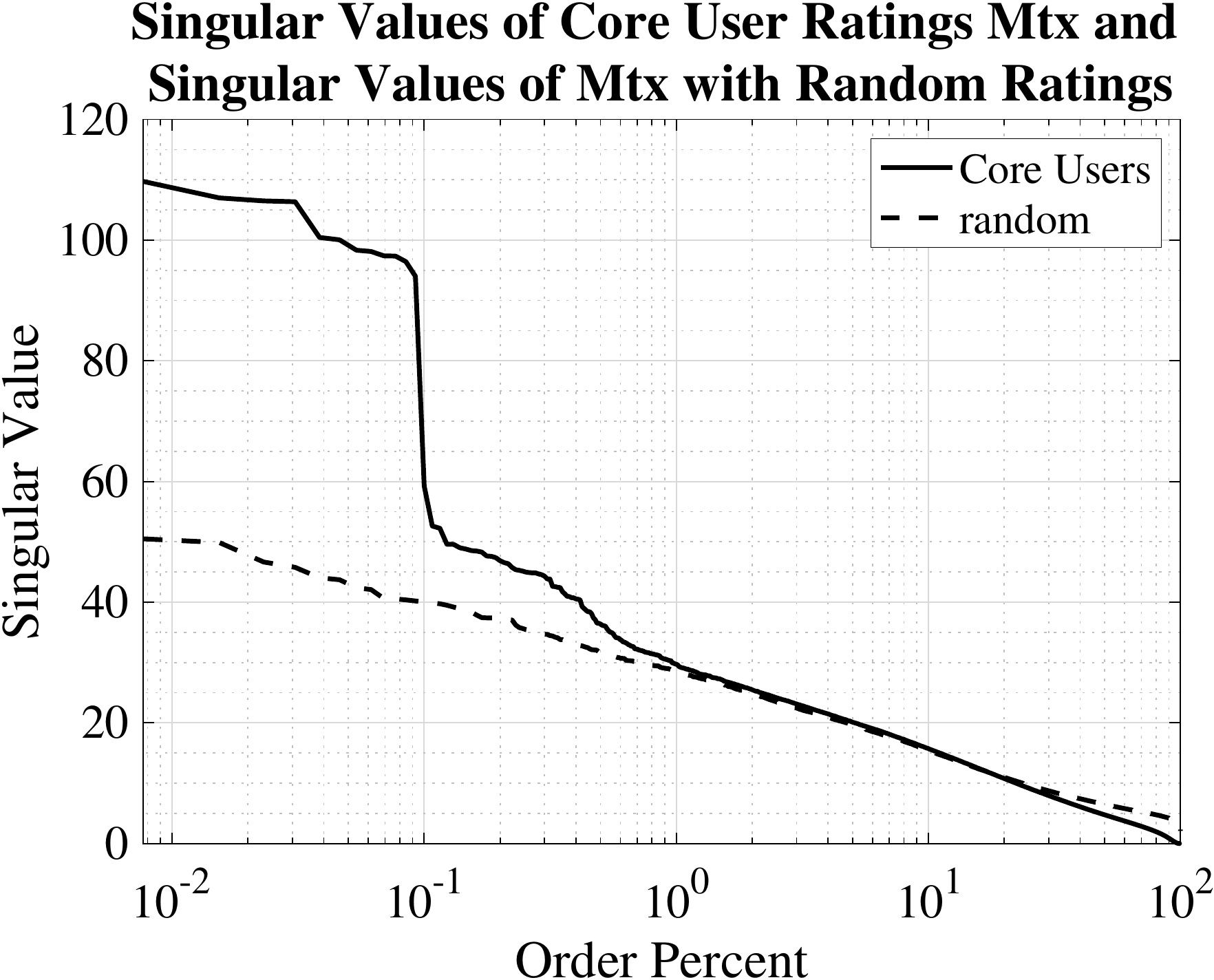}
  \vspace{-.15in}
\caption{Singular Values of 13000 Core User Ratings Matrix where Core Users are selected from the ml-20m dataset \citep{MovieLens} using the most competitive selection method: the method used in the second row of Table \ref{ItemVectorsTestingErroronCoreUsers}. Singular Values of an equally sparse matrix with random ratings as the non-zero values. There are 13000 singular values, where by convention lower order singular values have larger value.  The x-axis is labeled as the order percent, so the $i^{th}$ singular value would have $x$-tick value $i/130$. }\label{CUSVfig}
\end{minipage}
\hfill
\begin{minipage}[t]{.48\textwidth}
  \centering
  \includegraphics[width=\linewidth]{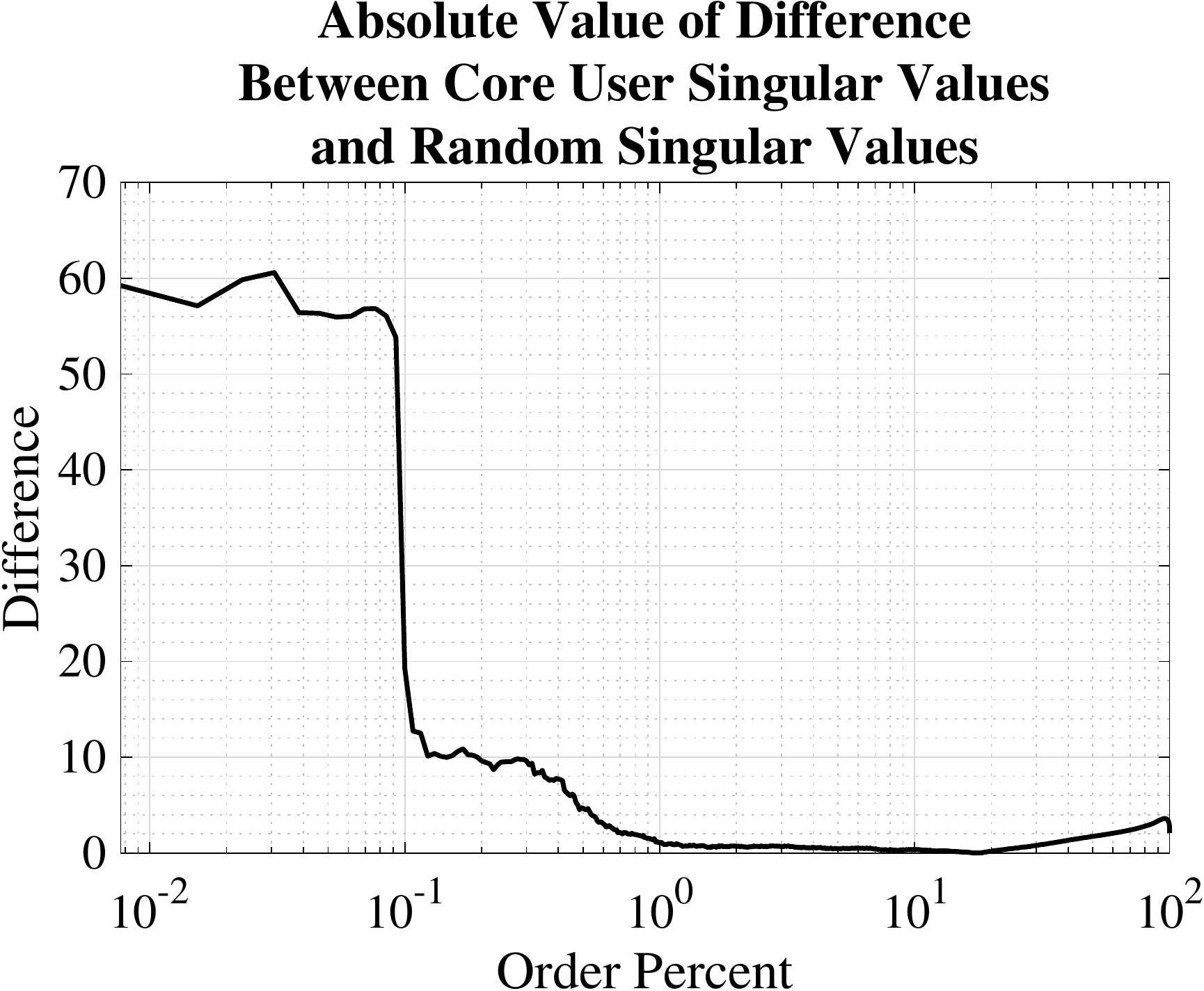}
  \vspace{-.15in}
  \caption{Absolute Value of the difference between the curves in Figure \ref{CUSVfig}.}\label{CUSVdifffig}
  \end{minipage}\hfill
\end{figure}

\XX{

\begin{figure}[h]
  \centering
  \includegraphics[width=0.5\linewidth]{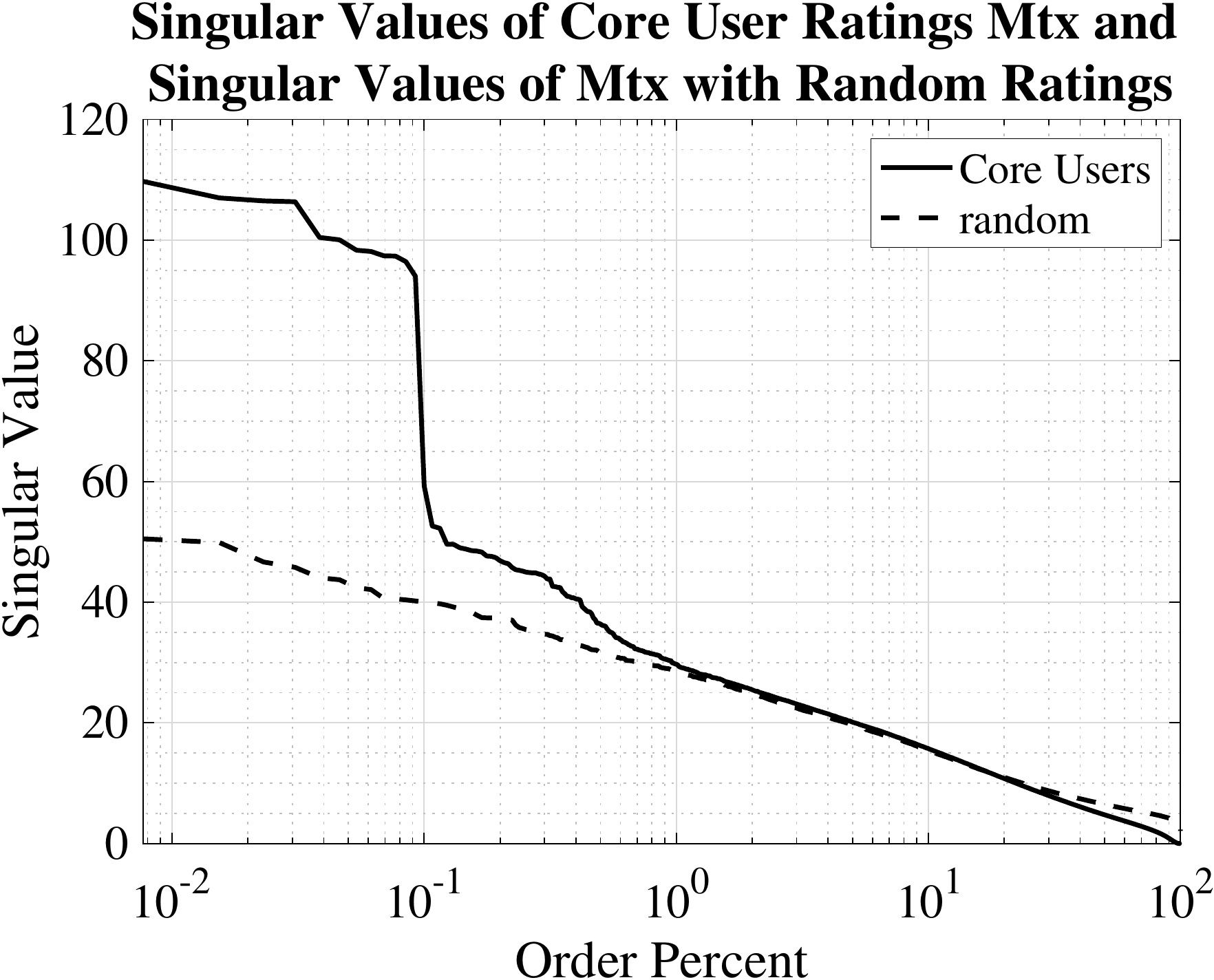}
\caption{Singular Values of 13000 Core User Ratings Matrix where Core Users are selected from the ml-20m dataset \citep{MovieLens} using the most competitive selection method: the method used in the second row of Table \ref{ItemVectorsTestingErroronCoreUsers}. Singular Values of an equally sparse matrix with random ratings as the non-zero values. There are 13000 singular values, where by convention lower order singular values have larger value.  The x-axis is labeled as the order percent, so the $i^{th}$ singular value would have $x$-tick value $i/130$. }\label{CUSVfig}
\end{figure}

\begin{figure}[h]
  \centering
  \includegraphics[width=0.5\linewidth]{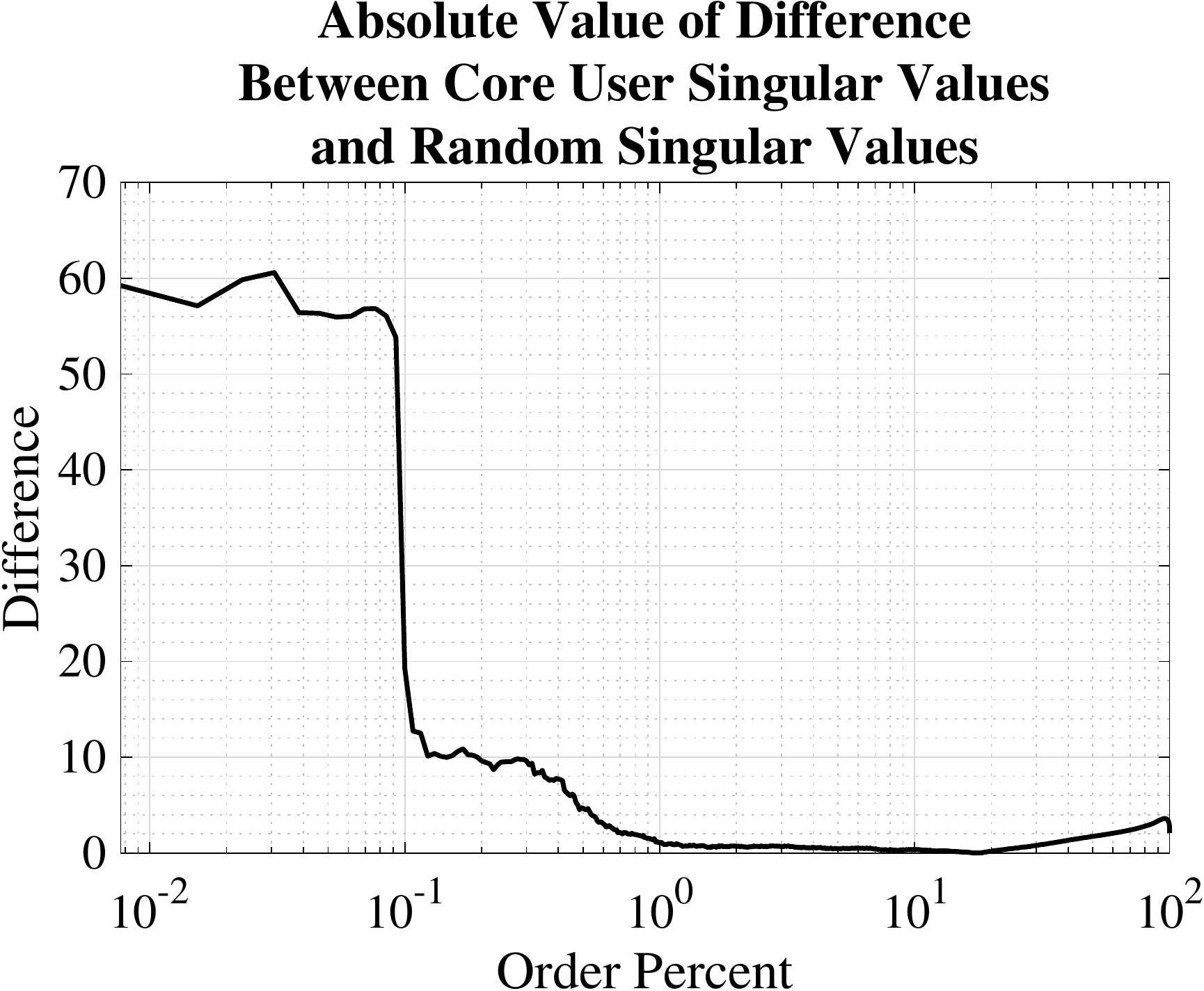}
\caption{Absolute Value of the difference between the curves in the Figure \ref{CUSVfig}.}\label{CUSVdifffig}
\end{figure}

}

\XX{

\begin{figure}
\centering
\begin{minipage}{.49\textwidth}
  \centering
  \includegraphics[width=0.5\linewidth]{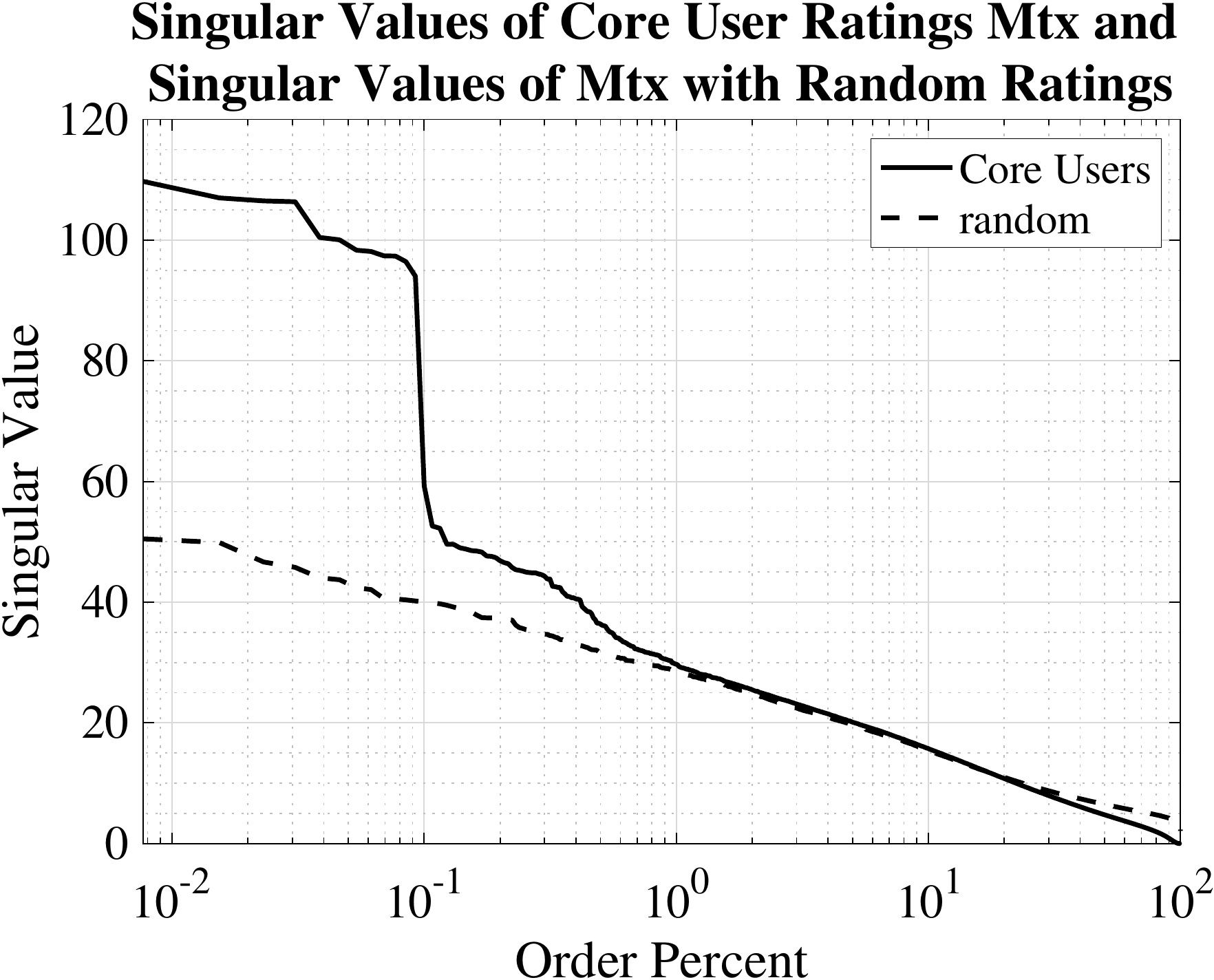}
\end{minipage}%
\hfill
\begin{minipage}{.49\textwidth}
  \centering
 \includegraphics[width=0.5\linewidth]{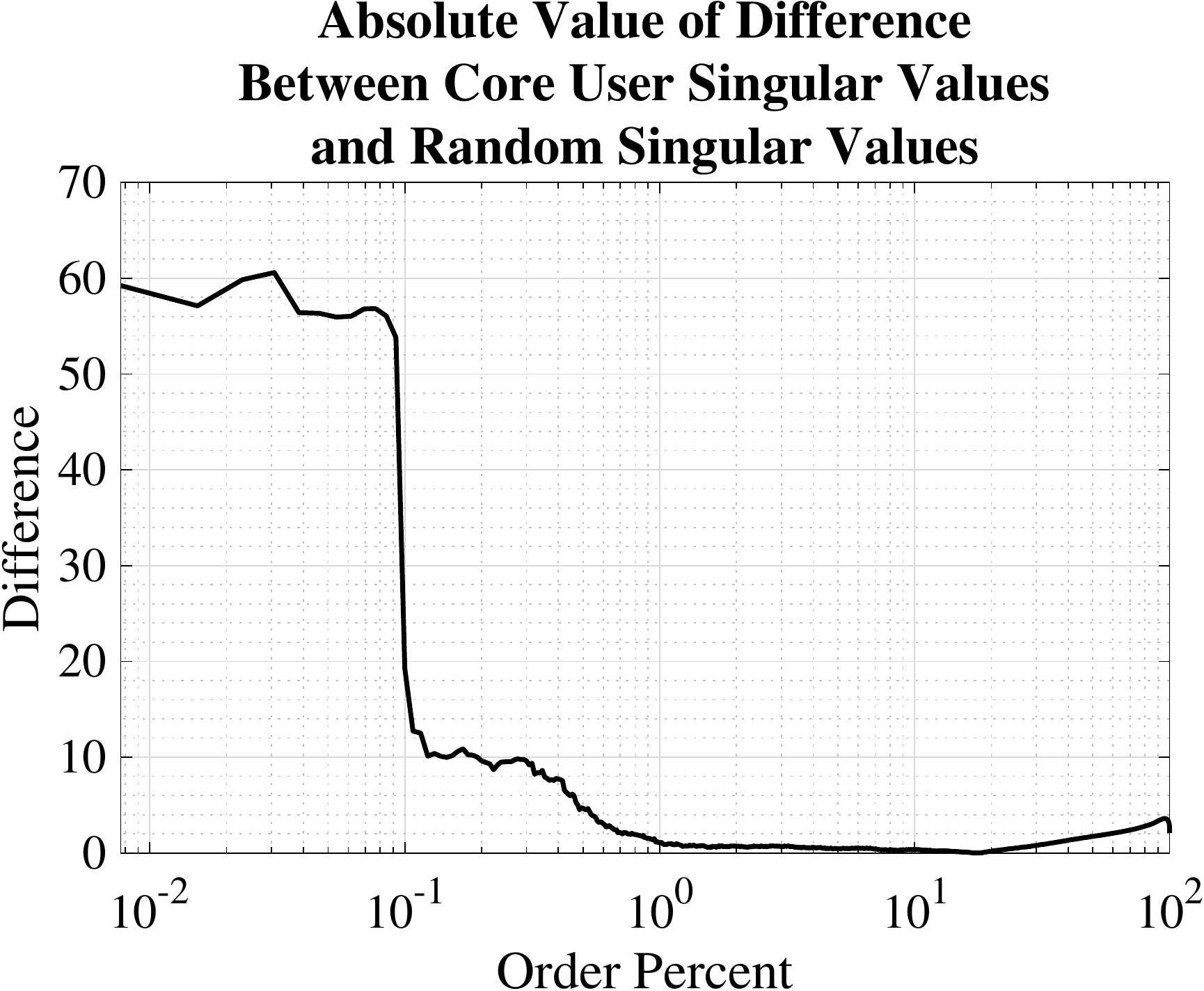}
\end{minipage}%
\caption{
(Left) Singular Values of 13000 Core User Ratings Matrix where Core Users are selected from the ml-20m dataset \citep{MovieLens} using the most competitive selection method: the method used in the second row of Table \ref{ItemVectorsTestingErroronCoreUsers}. Singular Values of an equally sparse matrix with random ratings as the non-zero values. There are 13000 singular values, where by convention lower order singular values have larger value.  The x-axis is labeled as the order percent, so the $i^{th}$ singular value would have $x$-tick value $i/130$. (Right) Absolute Value of the difference between the curves in the Left Figure.
}
\label{CUSVfig}\label{CUSVdifffig}
\end{figure}

}

\FloatBarrier


We now move to discussing the results of learning ACUs using Algorithm \ref{algGenerateACUs}. We initialized our ACUs are as Core Users selected with the most competitive selection method: the method used in the second row of Table \ref{ItemVectorsTestingErroronCoreUsers}. 

Figures \ref{IVTEfig} shows the item vectors testing error of 13000 ACUs over iterations, where an by an iteration of Algorithm \ref{algGenerateACUs} we are referring to one loop beginning on line \ref{outerwhile}. For each test, we averaged the results of Algorithm \ref{algItemVectorTestingError} over 20 runs where each run was given 200 independent testing users and either 2600 or 650 ACU item vectors.  As in our Core User tests, in each run we stopped Algorithm \ref{algItemVectorTestingError} when we reached a training error of 0.2. In both testing and training we used 13 row blocks, or 1000 ACUs per block. Whether we use 2600 or 650 ACU item vectors we can see clear improvement compared to the real Core User item vectors testing error, which is simply the $y-$intercept of these graphs. The best item vectors testing error using 650, or $5\%$ of the 13000 ACU item vectors is better than the item vectors testing error of 27000 Core Users ($20\%$ of the users in the complete dataset) when using using 650 Core User item vectors. The best item vectors testing error using 650 of the 13000 ACU item vectors is 0.987, while the item vectors testing error of 27000 Core Users when using using 650 Core User item vectors is 1.03.

Since the Core Users are stored in the same memory format as the original complete dataset, retaining only $20\%$ of the users as Core Users results in approximately a $80\%$ memory reduction. The memory reduction of the Artificial Core Users depends on the number of latent factors one decides to store. If one chooses to store only $5\%$ of the resulting latent factors, both user and item vectors, then this results in a $36\%$ reduction in memory compared to the original data set. If one chooses to store only $5\%$ of the the item vectors then this results in a $57\%$ reduction in memory compared to the original data set. Additionally, the improvement in the item vectors testing error when using only $5\%$ of the the item vectors in Algorithm \ref{algItemVectorTestingError} compared to using $20\%$ of the the item vectors as shown in Figures \ref{IVTEfig} suggests that  these extra vector may be more noisy than informative.

\XX{
\begin{figure}
  \centering
\includegraphics[width=0.5\linewidth]{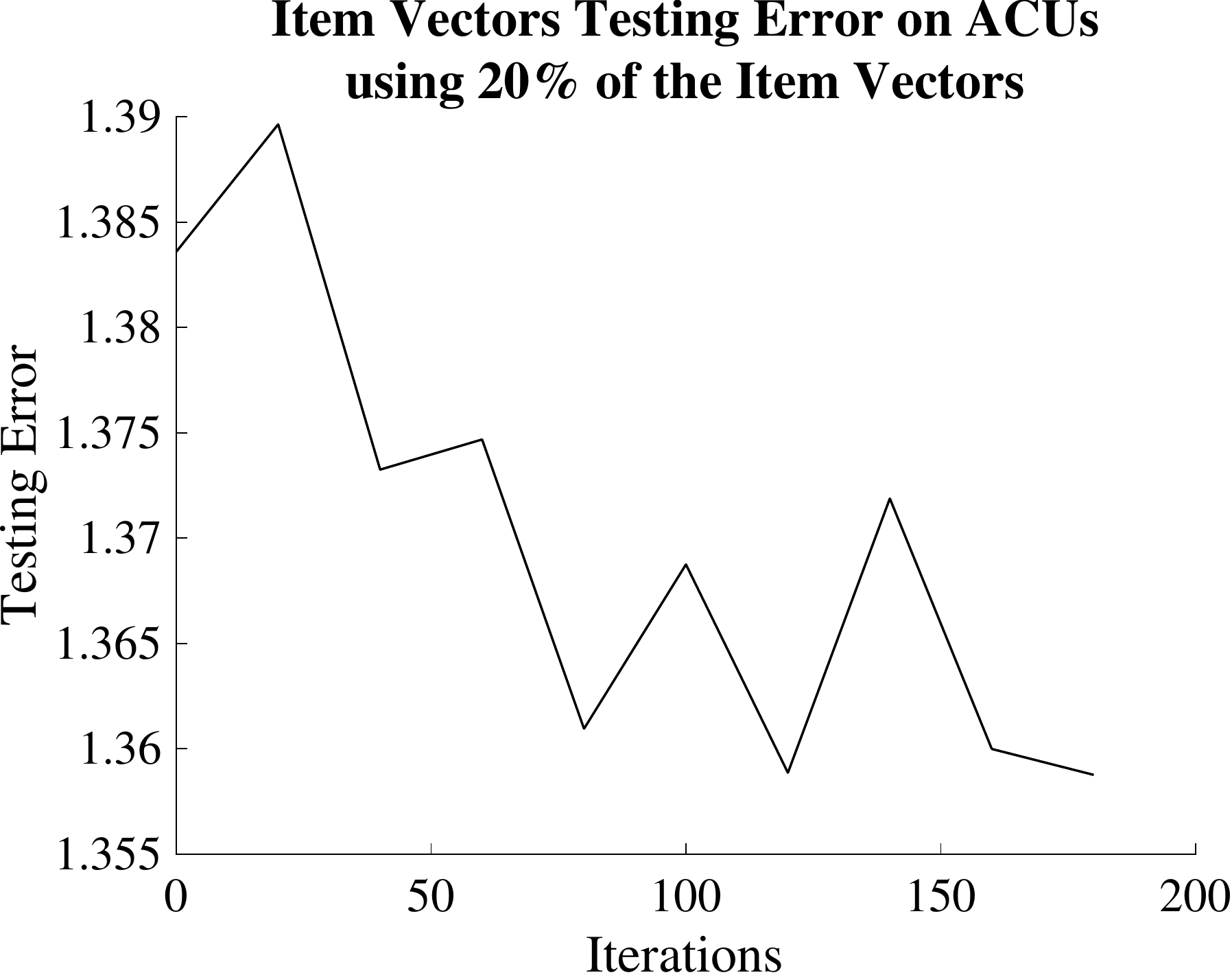}
\hfill\vspace{.01in}
\includegraphics[width=0.5\linewidth]{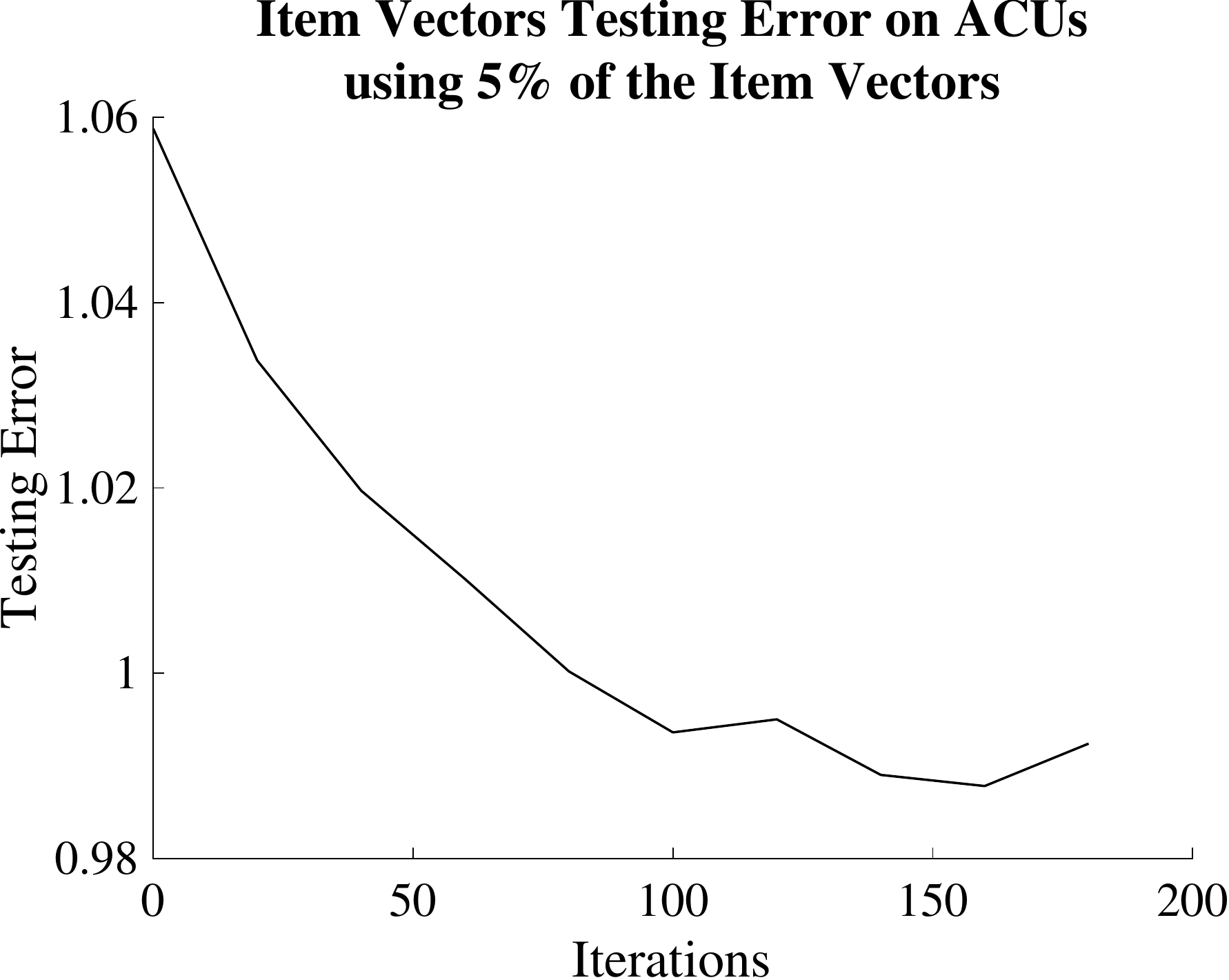}
  \caption{Item Vectors Testing Error of 13000 ACUs over iterations using the ml-20m dataset \citep{MovieLens}. For each test, we averaged the results of Algorithm \ref{algItemVectorTestingError} over 20 runs where each run was given 200 independent testing users and (Top) 2600/(Bottom) 650 ACU item vectors. In each run, we stopped Algorithm \ref{algItemVectorTestingError} when we reached a training error of 0.2.}\label{IVTEfig}
\end{figure}
}

\begin{figure}
\centering
\begin{minipage}{.49\textwidth}
  \centering
  \includegraphics[width=\linewidth]{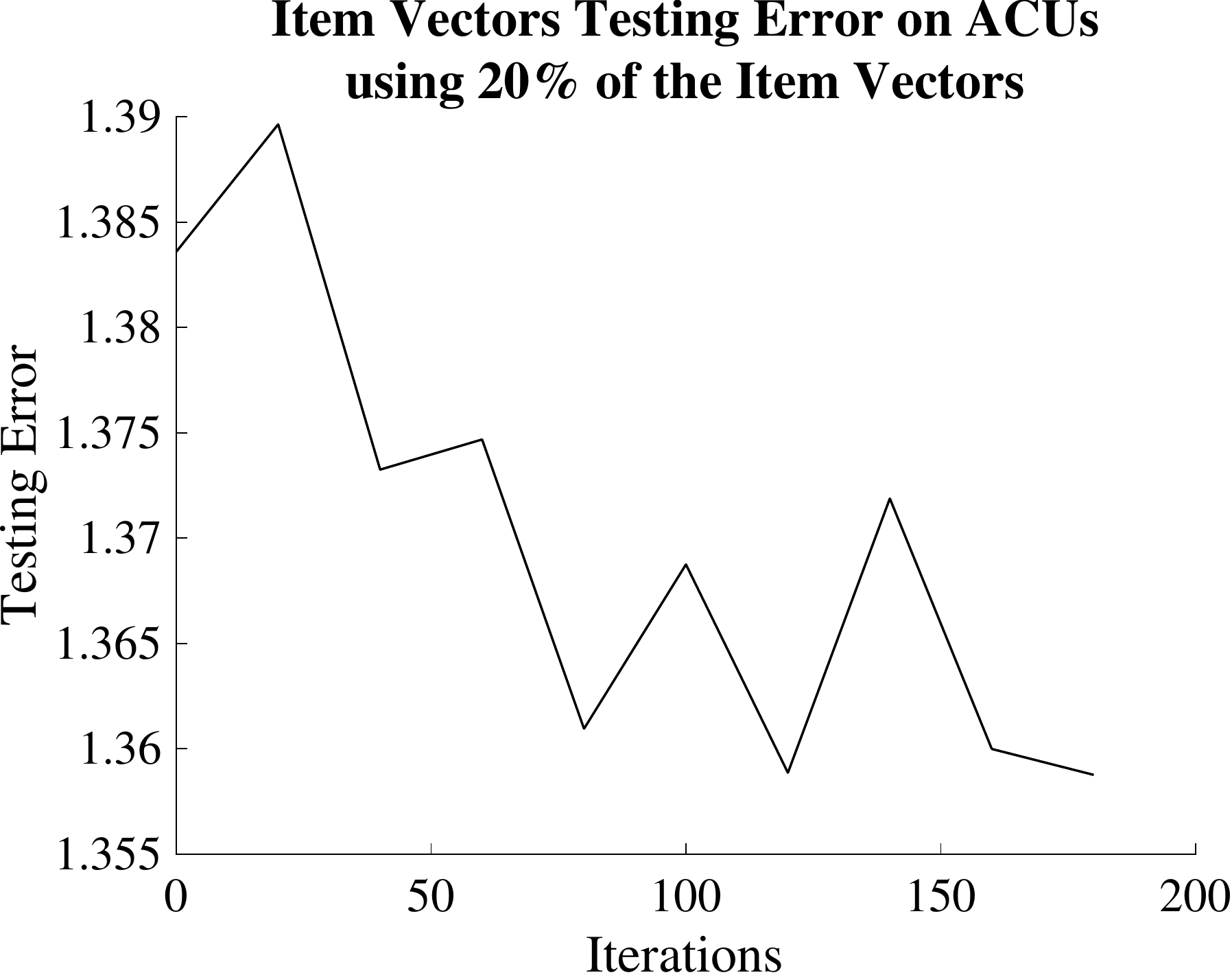}
\end{minipage}%
\hfill
\begin{minipage}{.49\textwidth}
  \centering
 \includegraphics[width=\linewidth]{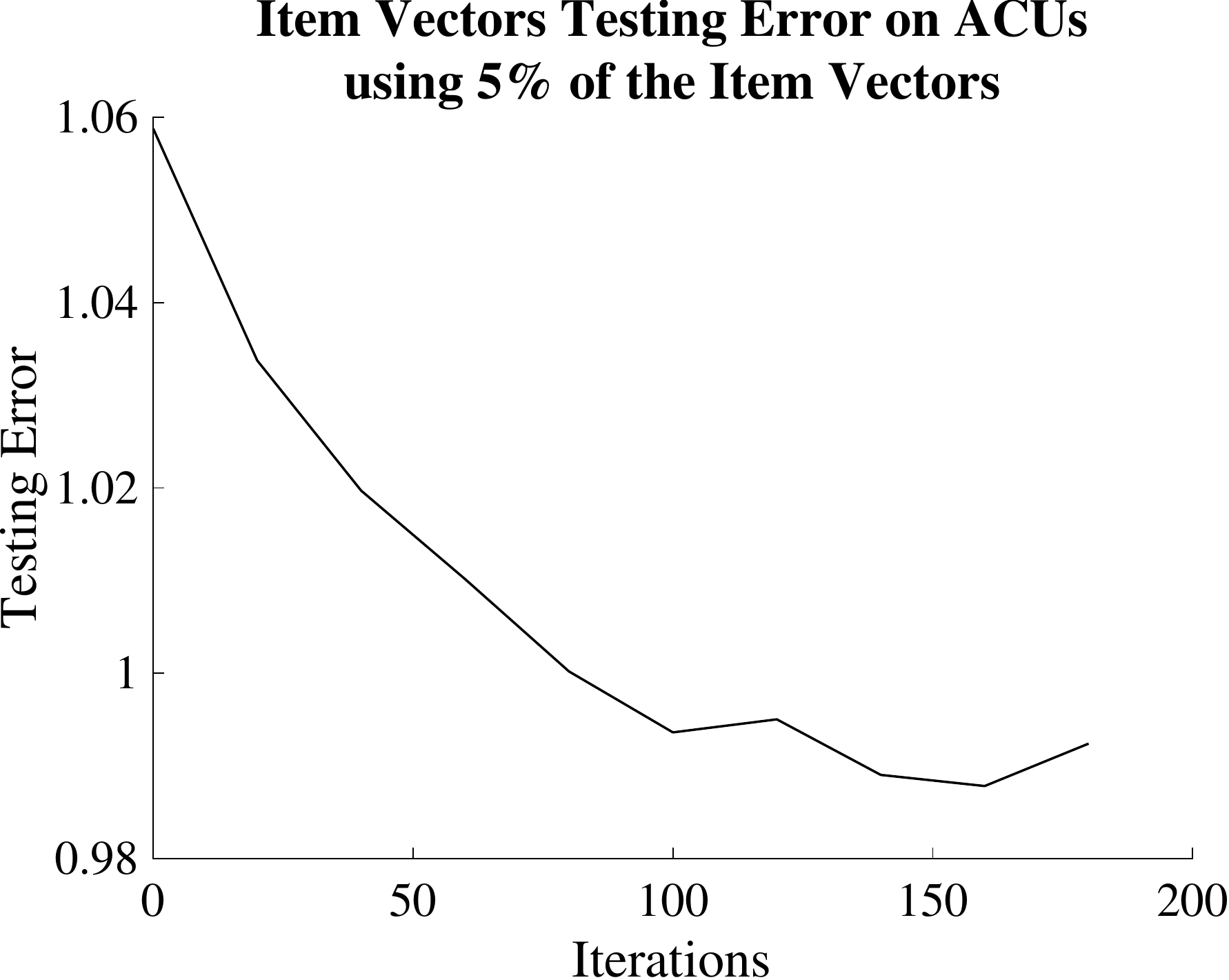}
\end{minipage}%
  \caption{Item Vectors Testing Error of 13000 ACUs over iterations using the ml-20m dataset \citep{MovieLens}. For each test, we averaged the results of Algorithm \ref{algItemVectorTestingError} over 20 runs where each run was given 200 independent testing users and (Left) 2600/(Right) 650 ACU item vectors. In each run, we stopped Algorithm \ref{algItemVectorTestingError} when we reached a training error of 0.2.}\label{IVTEfig}
\end{figure}

\FloatBarrier

Admittedly, Algorithm \ref{algGenerateACUs} learns relatively slowly. One loop beginning on line \ref{outerwhile} can take up to 4 minutes to complete.  We stopped running Algorithm \ref{algGenerateACUs} after 180 iterations at which point we reached an item vectors testing error of 1.36 with 2600 ACU item vectors and 0.99 with 650 ACU item vectors, which outperforms the most competitive real Core User methods. We also improved the sparse mean error \emph{slightly}, so our ACUs are a bit better centroids for the testing users than 13000 real Core Users are. Figure \ref{SMEfig} shows the sparse mean error of our ACUs over iterations calculated with Algorithm \ref{algSparseMeansError}. The second improvement is marginal, but demonstrates that our ACUs are still resemble an average group real users.


\XX{
\begin{figure}[h]
\vspace{-.2in}
  \centering
  \includegraphics[width=0.5\linewidth]{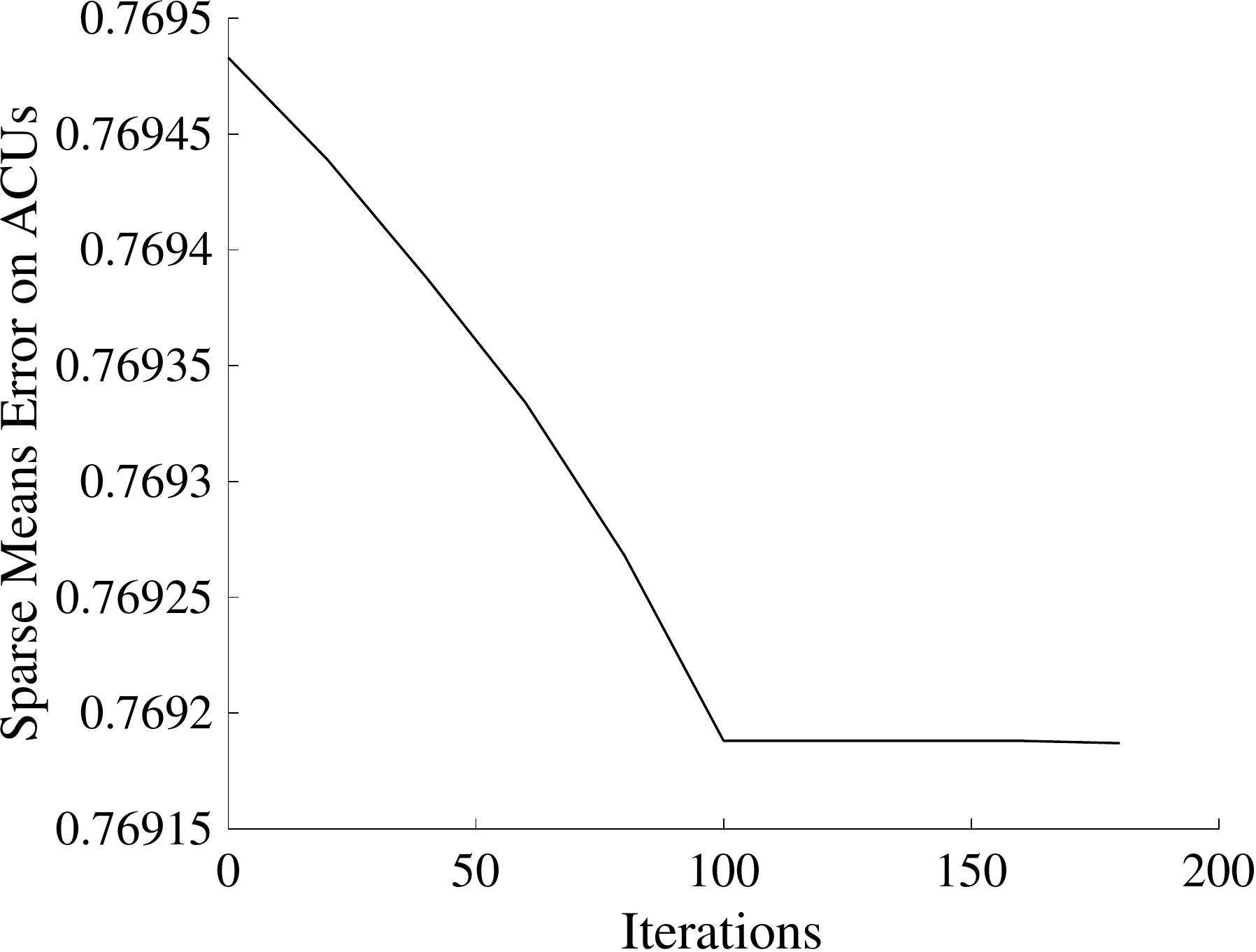}
  \caption{Sparse Mean Error of 13000 ACUs using the ml-20m dataset \citep{MovieLens}.}\label{SMEfig}
\hfill\vspace{.05in}
 \includegraphics[width=0.5\linewidth]{graphics/sv}
 \caption{650 largest singular values of 13000 trained ACU ratings matrix using the ml-20m dataset \citep{MovieLens}, compared to the largest singular values of the ratings matrix of 27000 Core Users ($20\%$ of the users in the complete dataset) and the largest singular values of the ratings matrix of 13000 Core Users ($10\%$ of the users in the complete dataset).}\label{recsysSVfig}
 \vspace{-.1in}
\end{figure}
}

\begin{figure}[h]
  \centering
  \includegraphics[width=0.49\linewidth]{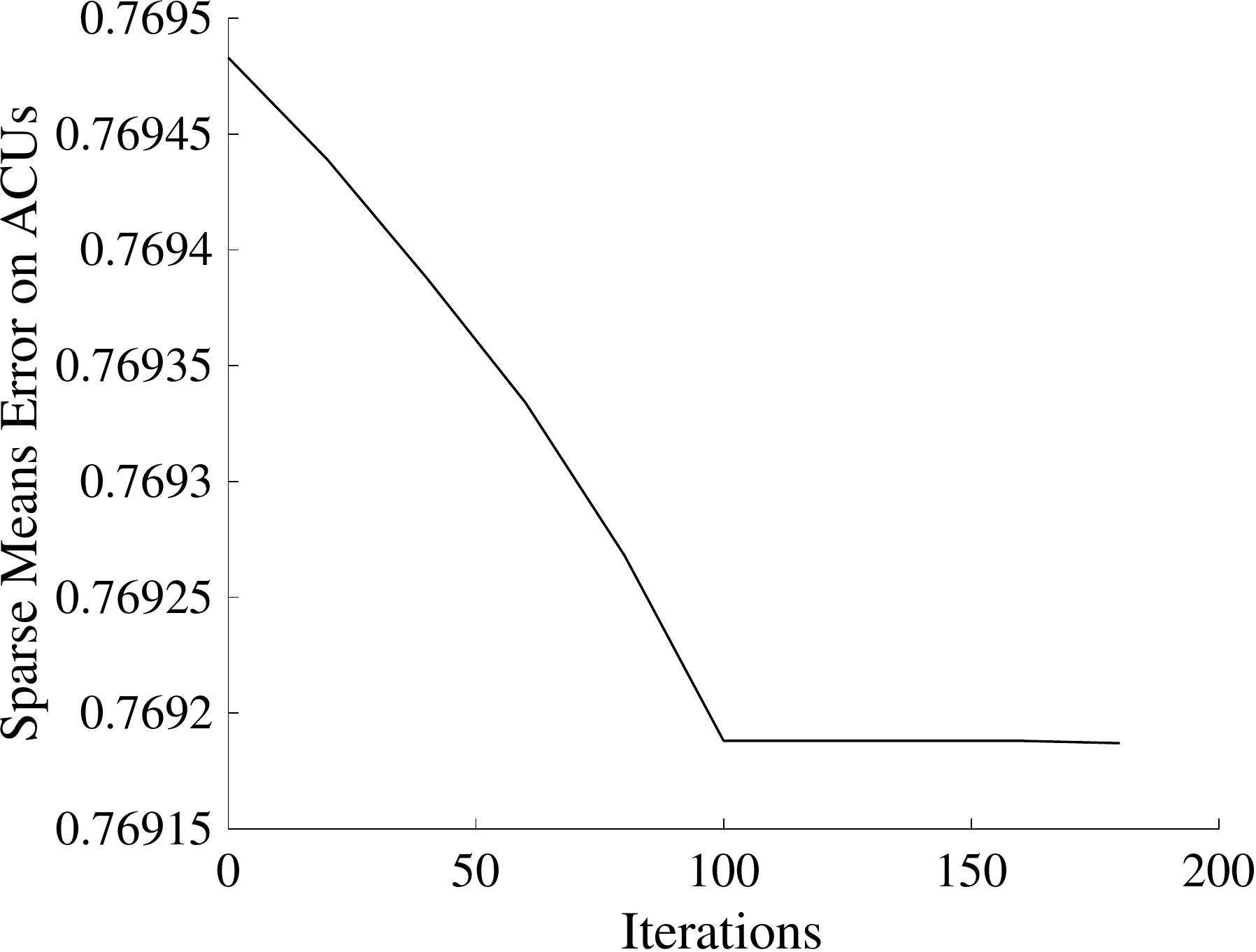}
  \caption{Sparse Mean Error of 13000 ACUs using the ml-20m dataset \citep{MovieLens}.}\label{SMEfig}
\end{figure}

We were able to condense the information of 27000 sparse Core Users into 13000 ACUs or approximately half the number of users. Since ACUs carry dense information, storing $5\%$ of the 13000 ACUs latent factors takes up about three times as much memory as 27000 sparse Core Users do, but $5\%$ of the 13000 ACUs latent factors achieves a smaller item vectors testing error than the same number of Core User latent vectors when using 27000 sparse Core Users. We can see from Figure \ref{recsysSVfig} that the largest singular values of the ACU ratings matrix, that had matched the largest singular values of the ratings matrix with 13000 Core Users at initialization, has shifted toward the the largest singular values of the ratings matrix with 27000 Core Users after 180 training iterations.

\begin{figure}[h]
  \centering
 \includegraphics[width=0.52\linewidth]{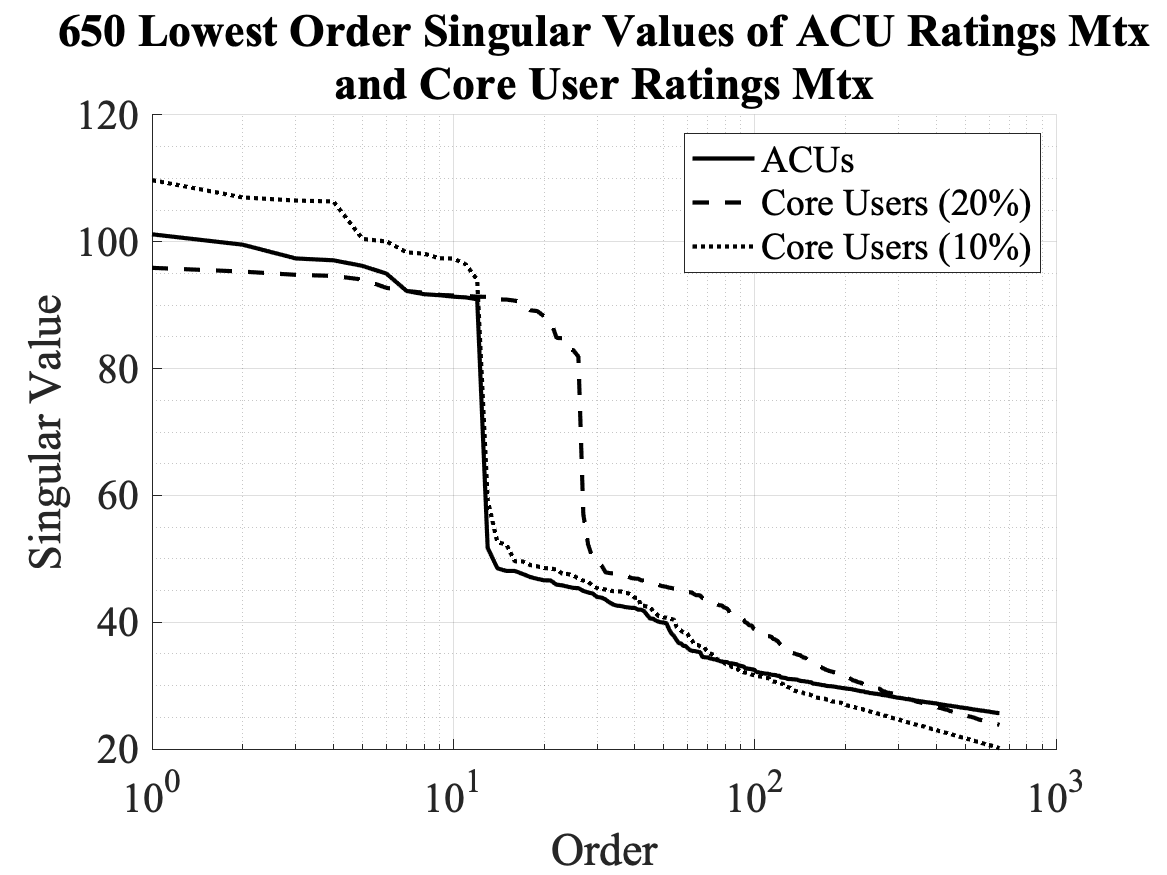}
 \caption{650 largest singular values of 13000 trained ACU ratings matrix using the ml-20m dataset \citep{MovieLens}, compared to the largest singular values of the ratings matrix of 27000 Core Users ($20\%$ of the users in the complete dataset) and the largest singular values of the ratings matrix of 13000 Core Users ($10\%$ of the users in the complete dataset).}\label{recsysSVfig}
\end{figure}

\XX{
\begin{figure}[h]
  \centering
  \begin{minipage}[t]{.48\textwidth}
   \centering
  \includegraphics[width=0.5\linewidth]{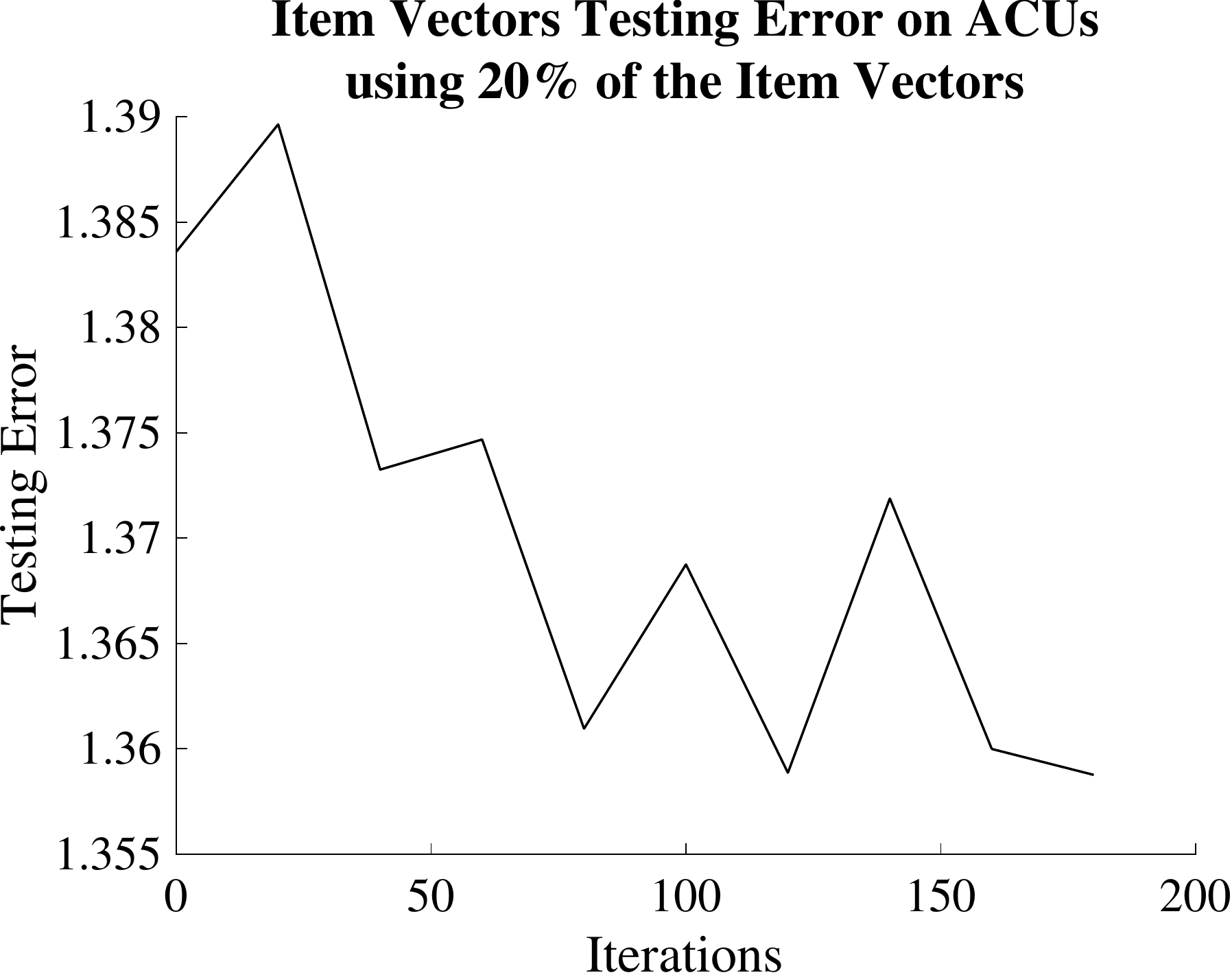}
  \caption{Item Vectors Testing Error of 13000 ACUs over iterations using the ml-20m dataset \citep{MovieLens}. For each test, we averaged the results of Algorithm \ref{algItemVectorTestingError} over 20 runs where each run was given 200 independent testing users and 2600 ACU item vectors. In each run, we stopped Algorithm \ref{algItemVectorTestingError} when we reached a training error of 0.2.}\label{IVTEfig}
\end{minipage}
\hfill
\begin{minipage}[t]{.48\textwidth}
  \centering
  \includegraphics[width=0.5\linewidth]{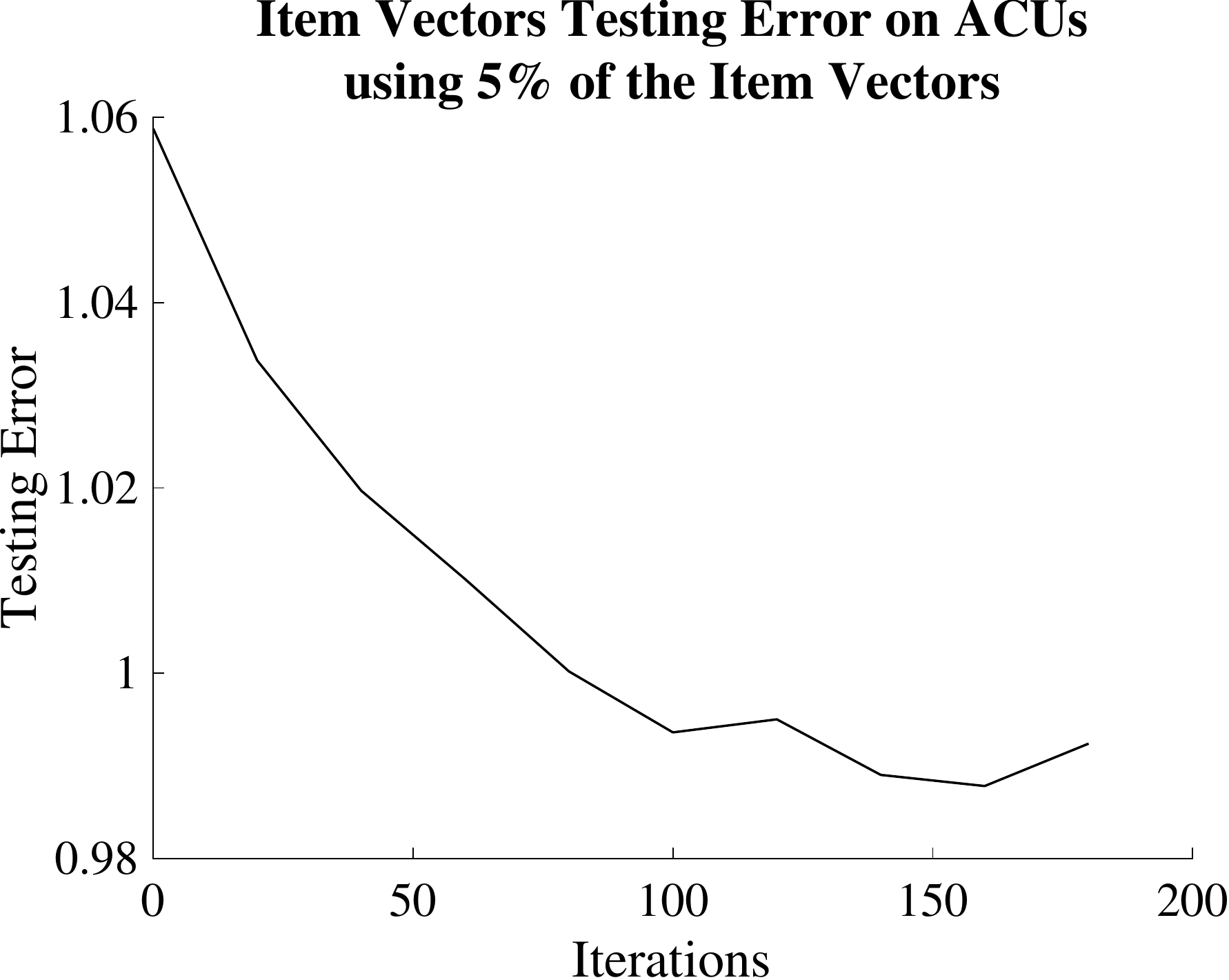}
  \caption{Item Vectors Testing Error of 13000 ACUs over iterations using the ml-20m dataset \citep{MovieLens}. For each test, we averaged the results of Algorithm \ref{algItemVectorTestingError} over 20 runs where each run was given 200 independent testing users and 650 ACU item vectors. In each run, we stopped Algorithm \ref{algItemVectorTestingError} when we reached a training error of 0.2.}
  \end{minipage}\hfill
\end{figure}
}

\XX{

\begin{table}
  \caption{Item Vectors Testing Error on 13000 Core Users using ml-20m \citep{MovieLens}.}
  \label{ItemVectorsTestingErroronCoreUsers}
  \begin{tabular}{ccccl}
    \toprule
    Item Similarity Used&Ratings Used&Frequency-Based&Training Set Error&Probe Set Error\\
    \midrule
    yes & yes& yes&0.20&2.50\\
   yes & yes& no&0.21&2.47\\
   yes & no& yes&0.18&3.14\\
   yes & no& no&0.18&3.09\\
     no & yes& yes&0.19&2.67\\
   no & yes& no&0.20&2.58\\
   no & no& yes&0.20&3.09\\
   no & no& no&0.20&3.00\\
  \bottomrule
\end{tabular}
\end{table}

\begin{table}
  \caption{Item Vectors Testing Error on 13000 Core Users using ml-20m \citep{MovieLens}.}
  \label{ItemVectorsTestingErroronCoreUsers}
  \begin{tabular}{ccccl}
    \toprule
    Item Similarity Used&Ratings Used&Frequency-Based&Training Set Error&Probe Set Error\\
    \midrule
    yes & yes& yes&0.45&3.37\\
   yes & yes& no&0.39&3.22\\
   yes & no& yes&0.32&4.54\\
   yes & no& no&0.39&4.50\\
     no & yes& yes&0.43&3.32\\
   no & yes& no&0.46&3.25\\
   no & no& yes&&\\
   no & no& no&&\\
  \bottomrule
\end{tabular}
\end{table}

\begin{table}
  \caption{Item Vectors Testing Error on 13000 Core Users using ml-20m \citep{MovieLens}.}
  \label{ItemVectorsTestingErroronCoreUsers}
  \begin{tabular}{ccccl}
    \toprule
    Item Similarity Used&Ratings Used&Frequency-Based&Training Error&Error on Probe Set\\
    \midrule
    yes & yes& yes&0.33&2.37\\
   yes & yes& no&0.38&2.36\\
   yes & no& yes&0.23&3.13\\
   yes & no& no&&\\
     no & yes& yes&&\\
   no & yes& no&&\\
   no & no& yes&&\\
   no & no& no&&\\
  \bottomrule
\end{tabular}
\end{table}

\begin{table}
  \caption{Item Vectors Testing Error on 13000 Core Users using ml-20m \citep{MovieLens}.}
  \label{ItemVectorsTestingErroronCoreUsers}
  \begin{tabular}{ccccl}
    \toprule
    Item Similarity Used&Ratings Used&Frequency-Based&Error\\
    \midrule
    yes & yes& yes&0.35&3.74\\
   yes & yes& no&0.41&3.61\\
   yes & no& yes&0.29&5.10\\
   yes & no& no&0.29&4.97\\
     no & yes& yes&0.40&3.75\\
   no & yes& no&&3.58\\
   no & no& yes&0.20&4.95\\
   no & no& no&0.27&4.79\\
  \bottomrule
\end{tabular}
\end{table}

}
\XX{
\subsection{Rent the runway}
\begin{table}
  \caption{Item Vectors Testing Error of 10000 Core Users collected with previously existing methods using the Rent the Runway dataset \citep{}). We averaged the results of Algorithm \ref{algItemVectorTestingError} over 75 runs where each run was given 200 independent testing users and 2600 of the Core User Item Vectors, or $\approx50\%\ $ of the singular value mass. In each run, we stopped Algorithm \ref{algItemVectorTestingError} when we reached a training error of 0.2. }
  \label{ItemVectorsTestingErroronCoreUsers}
  \begin{tabular}{ccccl}
    \toprule
    Item Similarity Used&Ratings Used&Frequency-Based&Probe Entry Set Error\\
    \midrule
   yes & yes& no&\bf{0.646782}\\
   no & yes& no&0.645973\\
  \bottomrule
\end{tabular}
\end{table}

}

\FloatBarrier

\section{Conclusion}

We have shown that our Artificial Core Users improve the recommendation accuracy of real Core Users while mimicking real user data. Because they act as good centroids for the complete dataset, they can be considered good representatives for real user clusters. They can be stored efficiently, yet they have dense ratings information which is more immediately interpretable than sparse data. We have removed the representation limits of Core Users and shown that an iterative training process can improve the recommendation accuracy of Core Users producing data that continues to resemble that of real users, conserve memory and improve recommendation efficiency.

  \XX{
Improving the recommendation accuracy of Core Users with ACUs makes data abstraction more viable across all applications.

  Core Users make out-of-sample recommendations more effective and efficient by reducing redundancy in the dataset, but Core User ratings remain sparse which is less informative to the naked eye. ACUs on the other hand provide dense interpretable data. Additionally, since ACUs have dense rating information, ACUs likely require less users than real Core Users to achieve the same recommendation accuracy.

  However, available Core User methods have limited representation ability, in that they are selected from existing user data. In other words, the recommendation success of Core Users is bounded by quality of user data available. 
Improving the recommendation accuracy of Core Users makes data abstraction more effective for applications in data augmentation, bots mimicking population behavior, data mining, privacy, statistics and many more. In this chapter, we develop a method of generating Artificial Core Users (ACUs) that improves the recommendation accuracy of real Core Users. 
We combine latent factor models, ensemble boosting and K-means clustering, to generate a small set of Artificial Core Users (ACUs) from real Core User data. Our ACUs incur a small amount of additional memory storage when compared to real Core Users, but remain a reduction in memory storage compared to the original dataset. Artificial Core Users improve the recommendation accuracy of real Core Users while remaining good centroids for the complete recommendation dataset. Since ACUs act as good centroids for the complete dataset, ACUs blend in with the real dataset even though they are generated artificially. But unlike real Core Users, ACUs have complete ratings on all items, providing more immediately interpretable information to scientists.
}

\XX{
\subsubsection*{Acknowledgements}
All acknowledgments go at the end of the paper, including thanks to reviewers who gave useful comments, to colleagues who contributed to the ideas, and to funding agencies and corporate sponsors that provided financial support. 
To preserve the anonymity, please include acknowledgments \emph{only} in the camera-ready papers.
}

\XX{


\bibliographystyle{plainnat}
 \bibliography{references}
}

\end{document}